\documentstyle[prb,aps,epsf,psfig]{revtex}

\begin{document}
\twocolumn[\hsize\textwidth\columnwidth\hsize\csname@twocolumnfalse\endcsname

\title{An asymptotical von-Neumann measurement strategy for
solid-state qubits}

\author {F.K.\ Wilhelm}

\address{Quantum Transport Group,  Technische Natuurkunde, TU Delft,
P.O. Box 5046, 2600 GA Delft, The Netherlands\\ and Sektion Physik and
CeNS, Ludwig-Maximilians-Universit\"at, Theresienstr.\ 37, 80333
M\"unchen, Germany}

\maketitle

\begin{abstract}
A  measurement on a macroscopic quantum system does in general not
lead to a projection of the wavefunction in the basis of the detector
as predicted by von-Neumann's postulate. Hence,  it is a question of
fundametal interest, how the preferred basis onto which the state is
projected is  selected out of the macroscopic Hilbert space of the
system.  Detector-dominated von-Neumann measurements are also
desirable for both quantum computation and verification of quantum
mechanics on a macroscopic scale. The connection of these
questions to the predictions of the spin-boson modelis outlined.  I propose a
measurement strategy,  which  uses the entanglement of the qubit with
a weakly damped harmonic oscillator.  It is shown, that the degree of
entanglement controls the degree of  renormalization of the qubit and
identify, that this is equivalent to the  degree to which the
measurement is detector-dominated.  This measurement
very rapidly decoheres the initial state, but the thermalization is
slow. The  implementation in Josephson quantum bits is described and
it is shown that this strategy also has practical advantages for the
experimental implementation.
\end{abstract}
 ]

The field of quantum computation\cite{ourfavoritereviews} has been
experimentally pioneered in quantum optics, atomic physics and nuclear
magnetic resonance (NMR). In these quantum-mechanical systems with few
degrees of freedom and strong quantum coherence,  the measurement
devices  (``meters'') are well described and can be classified into
two types: In atomic physics, e.g., ``strong'' measurements can be
performed,  which satisfy  von Neumann's measurement postulate
\cite{vonNeumann}, i.e. the state of the system is projected onto the
eigenstate of the {\em meter} corresponding to the measurement result.
In NMR, on the other hand, the meter couples weakly to  each
individual spin and decoheres it only weakly. In order to still obtain
enough signal and information, the measurement is performed on an
ensemble of qubits.

These qubits cannot be easily integrated to large-scale circuits.
Thus, solid-state qubits, which can be lithographically manufactured,
are a promising alternative. Solid state systems  consist of many
degrees of freedom, hence quantum coherence can so far  only be
maintained over very short times \cite{Nakamura,Caspar}.  It was
proposed that superconducting Josephson circuits in the charge
\cite{Nakamura,Makhlin} or flux  \cite{Caspar,HansScience} regime
could act as solid state qubits with appreciable coherence times.  In
these cases, the measurement apparatus is permanently close to the
qubit, although the interaction may effectively be switched off
\cite{Shnirman,CasparLong}.  The measurement process in this system
can be described within the spin-boson \cite{Leggett,Milena} or
related models \cite{Shnirman,Korotkov,Averin}.

From a density-matrix description, we can obtain detailed (although
incomplete) information about the dynamics of the measurement:  After
a dephasing time $\tau_{\rm\phi}$, the density
matrix is brought into an incoherent mixture and after the relaxation time
$\tau_{\rm r}$ it  thermalizes and the information about
the initial state  is lost \cite{Shnirman}.  In order to render
$\tau_{\rm r}$ long enough, usually
\cite{Nakamura,Caspar,Makhlin}  the meter is only weakly coupled to
the qubit.  This makes it necessary to ensemble-average by
repeating the measurement. Theoretical research
\cite{Shnirman,Korotkov,Averin} shows, that an optimization of these
weak measurements allows for single-shot measurements without
averaging, by waiting longer that the dephasing time. These are {\em
optimized weak measurements} or {\em qubit dominated} measurements: 
They completely decohere the state of the
qubit, however, the final state is {\em not} an eigenstate of the
measured observable, but of the qubit. 
Qubit and apparatus do
{\em not} get strongly entangled.  It has also  been shown
theoretically \cite{Shnirman},  that detector-dominated {\em strong} 
measurements
of superconducting qubits are possible, on the expense of $\tau_{\rm
R}$ being  very short, which sets a strong experimental challenge.  It
is a fundamental question, under which conditions a measurement
performed on a potentially macroscopic object follows the postulates
of quantum mechanics and how, in general, the preferred observable
basis is selected out of the large Hilbert space of the system and the
detector \cite{Zurek}. This question should be addressed using
specific models which describe actual detectors.  Moreover, there are
practical issues: i)  The theoretical signal to noise ratio of a weak
measurement is limited to  4 \cite{Averin} ii) Efficient quantum
algorithms like error correction \cite{error} or the test of Bell-type
inequalities \cite{LeggettGarg} rely on strong measurements.

In this paper, I am going to connect the abstract notions of quantum
measurement theory to the concepts of the spin-boson model, in
particular the issue of entanglement will be connected to scaling of
the tunnel matrix element.  I will outline a method how to perform
genuine detector dominated measurements in this context.

For definiteness, it is assumed that the variable of the quantum bit
which is measured is described by the pseudospin-operator
$\hat{\sigma}_{\rm z}$. When the measurement apparatus is coupled to
the qubit, the same term experiences a  fluctuating force,  which is
assumed to be Gaussian and  be modeled by a  bath of harmonic
oscillators. Consequently, we end up with the  spin-boson Hamiltonian
\cite{Leggett,WeissBuch}. After integrating out high frequencies,  its
pseudospin part reads
\begin{equation}
\hat{H}_{\rm eff}=\hbar\left(\frac{\epsilon}{2}\hat{\sigma}_z+\frac{\Delta_{\rm
eff}}{2}\hat{\sigma}_x\right)
\label{Heff}
\end{equation}
where the off-diagonal term $\Delta_{\rm eff}$ is in general rescaled
due to the environment as compared to the original splitting $\Delta$
of an isolated qubit.  The spin-boson model generally predicts
\cite{WeissBuch} the dynamics described in the previous section. In
particular,   after the dephasing time,  the density matrix onto a
mixture of {\em eigenstates of $H_{\rm eff}$}.  Usually, in the weak
coupling regime\cite{Leggett}, $\Delta_{\rm eff}$ is close to  the
bare $\Delta$ of the qubit and consequently the eigenstates of $H_{\rm
eff}$ are far from being eigenstates of  $\hat{\sigma}_{\rm
z}$. Consequently, the state of the qubit will not be procjected onto
the measured variable, i.e.\ the measurement is qubit-dominated.  A
detector-dominated measurement would be realized for  $\Delta_{\rm
eff}=0$\cite{Zimanyi}, i.e. when   $\hat{H}_{\rm
eff}=(\epsilon/2)\hat{\sigma_{\rm z}}$ and hence commutes with the
coupling to the meter.  A number of schemes allow to directly
suppress $\Delta$ using an external control parameter
\cite{Makhlin,HansScience,Orlando}. In what follows, I want to
describe, using a generic model, how this is accomplished by the
measurement apparatus itself, in agreement with the usual
understanding of quantum measurements.

Consider a qubit coupled to a single (big) harmonic oscillator, which
experiences linear friction, which is in turn described
quantum-mechanically through a bath of oscillators. The Hamiltonian of
this system reads
\begin{eqnarray}
\hat{H}&=&\hbar\left(\frac{\epsilon}{2}\hat{\sigma}_z+\frac{\Delta}{2}\hat{\sigma}_x\right)
+\frac{\hat{P}^2}{2M}
+\frac{M}{2}\Omega^2(\hat{X}-q\hat{\sigma}_z)^2\nonumber\\
&&+\sum_i\left(\frac{\hat{p}_i^2}{2m_i}
+\frac{m_i}{2}\omega_i^2(\hat{x}_i-(c_i/m_i\omega_i^2)\hat{X})^2\right)
\end{eqnarray}
where the dispacement $q$ characterizes the coupling of the qubit to
the  big oscillator.  The oscillator bath is characterized through an
ohmic spectral density $ J(\omega)=\sum \frac{\pi
c_i^2}{2m_i\omega_i}\delta(\omega-\omega_i) =M\Gamma\omega, $ where,
$\Gamma$ is the friction coefficient of the damped big oscillator.  It
was shown \cite{Garg}, that this system is equivalent to the
spin-boson model, with an effective spectral density
\begin{equation}
J_{\rm eff}(\omega)=2\pi\alpha\omega
\frac{\Omega^4}{(\omega^2-\Omega^2)^2+4\Gamma^2\omega^2},
\label{spectrum}
\end{equation}
where $\alpha=2Mq^2\Gamma/h$  is a dimensionless dissipation
coefficient, which here is assumed to be  small, $\alpha\ll1$.  From
now on, we want to concentrate on the case $\epsilon,\Delta\ll\Omega$.

\begin{figure}[htb]
\begin{minipage}{0.4\columnwidth}
\centerline{\psfig{figure=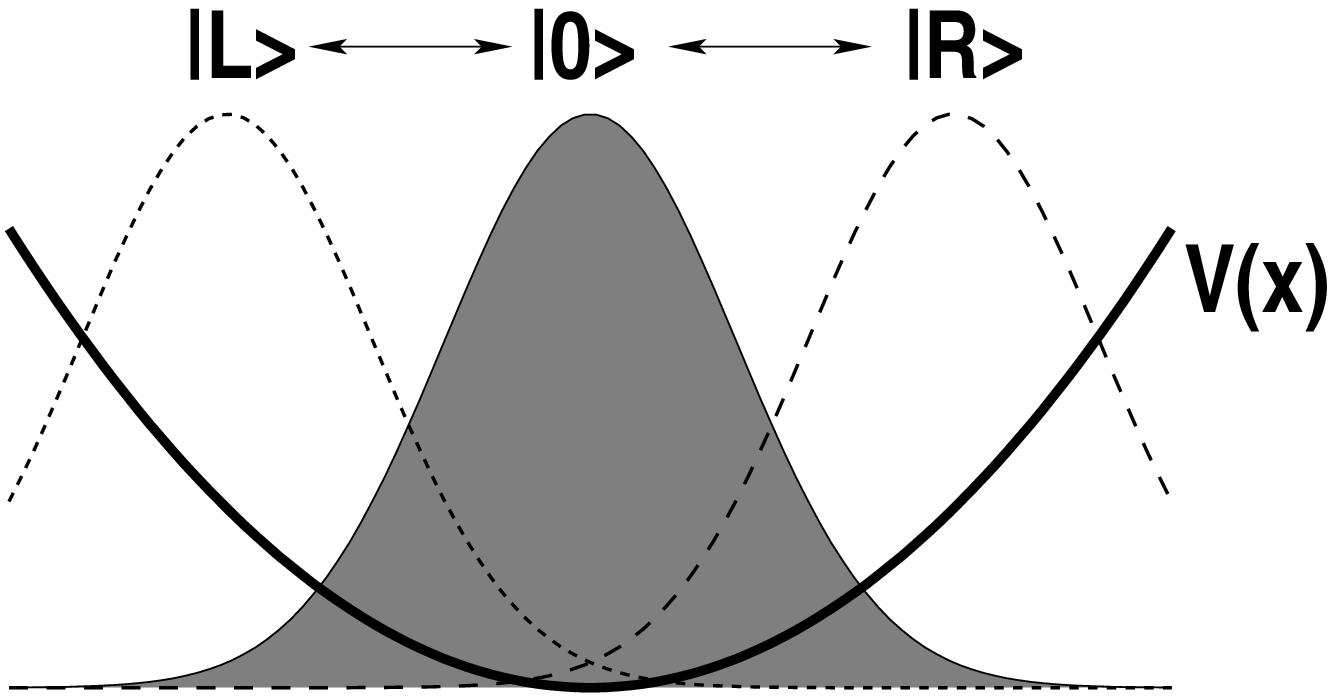,width=0.95\columnwidth}}
\end{minipage}
\begin{minipage}{0.55\columnwidth}
\centerline{\psfig{figure=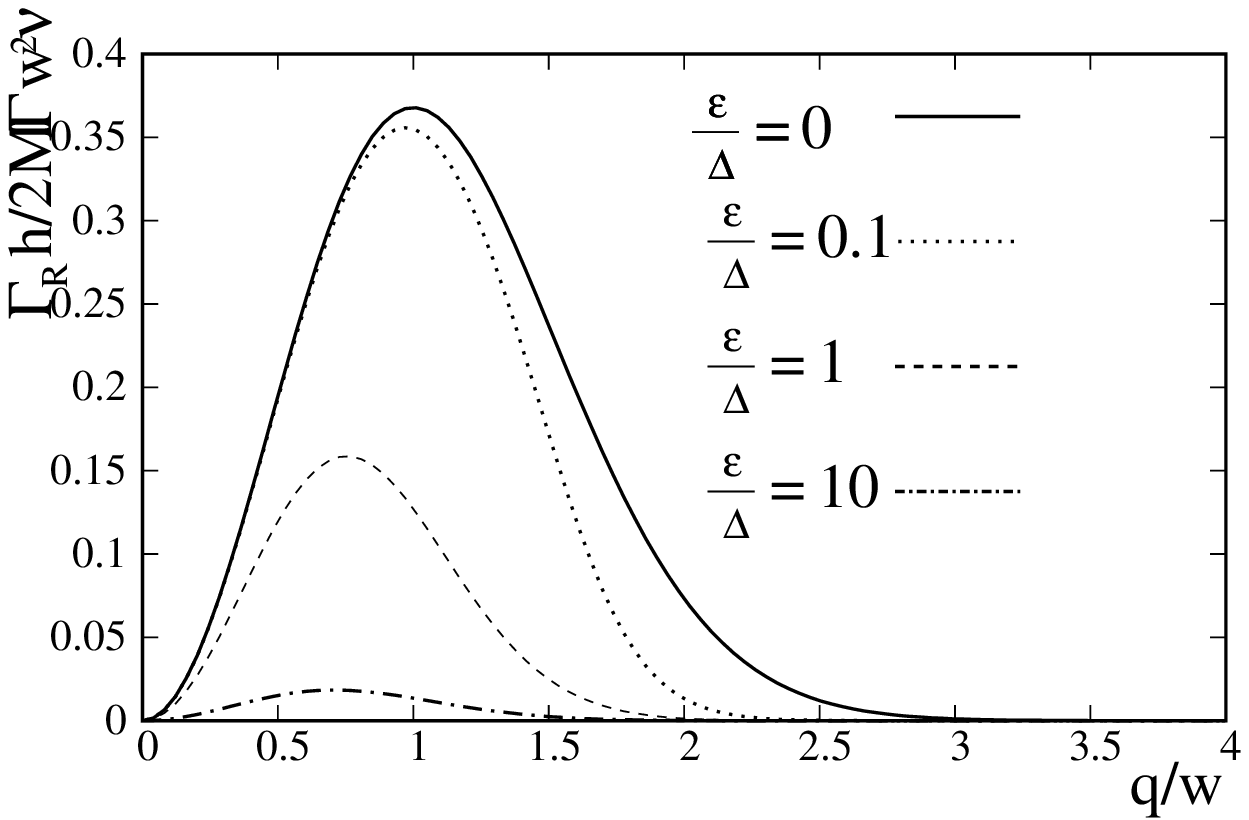,width=0.95\columnwidth}}
\end{minipage}
\caption{left: Visualization of the ground state $|0\rangle$ and the
coherent pointer-states $|L\rangle$ and $|R\rangle$ of the oscillator
\label{pointers} in the potential $V(x)$; right: Relaxation rates as a
function  of the coupling $q/w$ for different energy biases. $w$ is
the width of the ground state wavefunction of the pointer, $w=\sqrt{\hbar/M\Omega}$}
\end{figure}

At $\Gamma=0$, the low-energy Hilbert space is is spanned by
$|\pm\rangle_{\rm eff}=|\pm\rangle|L/R\rangle$ where $|\pm\rangle$ are
the basis states of the qubit,  $\sigma_z|\pm\rangle=\pm|\pm\rangle$
and $|L/R\rangle$ are coherent states of the harmonic oscillator
centered around $X=\pm q$, see Fig.\ \ref{pointers}, left.  So in a
general low-energy state $|\psi\rangle=a|+\rangle_{\rm
eff}+b|-\rangle_{\rm eff}$, $|a|^2+|b|^2=1$  qubit and oscillator are
entangled \cite{FN} and the oscillator states are pointers onto the
qubit states \cite{Zurek}.  In this low-energy basis, the Hamiltonian
acquires the form eq. (\ref{Heff}), with $\Delta_{\rm
eff}=\Delta\langle L|R\rangle=\Delta e^{-\eta}$ with $\eta=M\Omega
q^2/\hbar$.  Under an appropriate choice of parameters, we can achieve
$\eta>1$ and  $\Delta_{\rm eff}\ll\Delta$.  This choice corresponds to
the condition of almost  (meaning here and  henceforth ``up to an
error of $O(e^{-\eta})$'') {\em orthogonal} states in the environment,
which has been identified as the condition for an ideal
von-Neumann-measurement\cite{Zurek,Peres}.

For finite $\Gamma$,  this system can be analyzed  using adiabatic
renormalization \cite{Leggett,Chakravaty}. One finds \cite{note}
$\Delta_{\rm eff,damp} (\alpha)=\Delta
e^{-\frac{\eta}{1-\alpha}}(\Delta/\Omega)^{\frac{\alpha}{1-\alpha}}$ .
Thus,  finite dissipation $0<\alpha\ll1$ scales $\Delta$ down even
slightly further.

The coherence properties of our system can at
$\epsilon,\Delta,T\ll\Omega$ be studied using a systematic weak
damping  approximation \cite{WeissBuch} of the spin-boson model.  The
relaxation and dephasing rates, $\Gamma_{\rm r/\phi}=\tau_{\rm
r/\phi}^{-1}$  are given by
\begin{equation}
\Gamma_{\rm r}=\pi\alpha\frac{\Delta_{\rm eff}^2}{\nu_{\rm eff}}{\rm
coth} \left(\frac{\nu}{2T}\right)\;\; \Gamma_{\rm
\phi}=\frac{\Gamma_{\rm r}}{2} +2\pi\alpha k_{\rm B}\frac{\epsilon^2}{\nu_{\rm
eff}} T/\hbar
\label{gphi}
\end{equation}
where $\nu_{\rm eff}=\sqrt{\Delta_{\rm eff}^2+\epsilon^2}$.  In our
case, if $\eta>1$, $\Delta_{\rm eff}$ is
exponentially reduced compared to $\Delta$, transitions  between the
basis states are suppressed leaving relaxation very slow, i.e.\ 
the state
becomes almost localized or ``frozen'', see fig.\ \ref{pointers},
right.  The second contribution to $\Gamma_{\rm\phi}$ in eq.\
(\ref{gphi})  reflects dephasing processes which do not change the
qubit energy and are consequently not frozen.

The use of a weak damping approximation for $\Gamma_{\rm r,\phi}$ is
appropriate,  although $J(\omega)$ can be large at the peak and in
fact the down-scaling of $\Delta_{\rm eff}$ is  essentially a
nonperturbative effect. However, decoherence is mostly probing the
$J_{\rm eff}(\omega)$ around $\omega=\nu_{\rm eff}\ll\Omega$,  where
the weak damping condition holds.  This is supported by two
observations: i) if we project the full Hamiltonian onto its
low-energy Hilbert space spanned by $|\pm\rangle_{\rm eff}$, we find
an effective ohmic model leading to eq.\ \ref{gphi}.  ii) a full
nonperturbative  calculation \cite{note} based on the noninteracting
blip approximation (NIBA)  \cite{Leggett} reproduces both the scaling
and  $\Gamma_{\rm r}$  within the known \cite{WeissBuch} limitations
of NIBA.

The measurement can now be performed as follows: As a first step,  $q$
is adiabatically ramped from $q=0$  to a finite  $q_0$  where $\eta>1$
and $\Delta\rightarrow\Delta_{\rm eff}\ll\Delta$.  The adiabatic
theorem  predicts, that the state of the system evolves as
$(\alpha|+\rangle+\beta|-\rangle)\otimes |0\rangle \rightarrow
(\alpha^{\rm eff}|+\rangle|L\rangle+\beta^{\rm
eff}|-\rangle|R\rangle)$, where
\begin{equation}
\left(\matrix{\alpha^{\rm eff}\cr \beta^{\rm eff}\cr}\right)
=\left(\matrix{\cos\left(\frac{\theta^{\rm eff}-\theta}{2}\right)
&\sin\left(\frac{\theta^{\rm eff}-\theta}{2}\right)\cr
-\sin\left(\frac{\theta^{\rm
eff}-\theta}{2}\right)&\cos\left(\frac{\theta^{\rm
eff}-\theta}{2}\right)\cr}\right)\left(\matrix{\alpha\cr
\beta\cr}\right)
\end{equation}
and $\tan\theta^{\rm (eff)}=\epsilon/\Delta^{\rm (eff)}$.  The
condition for adiabaticity is $dq/dt\ll \nu_{\rm
eff}^2/2qM\Omega\Delta_{\rm eff}$,  i.e. for small $q$, the ramping
can be very fast.

When $\eta>1$,  the matrix element is scaled down and the state is
``pre-measured'' by  entanglement with well-separated pointer states
$L$ and $R$\cite{Zurek}.  Only now, we start the measurement, by
coupling the oscillator to the meter and decohering the  state is
projected onto the eigenstates of $\hat{H}_{\rm eff}$, which are close
to the ones of  $\hat{\sigma}_{\rm z}$. We can then read off the
position of the big oscillator serving as a pointer and  switch off
the meter (or $q$) again way before $\tau_{\rm r}$ without destroying
information by relaxation.

In practice, it will usually not be possible to switch the coupling
between oscillator and meter separately. Thus, before the entanglement
is established, the relaxation rate eq.\ \ref{gphi}  does {\em not}
profit from  the reduction of $\Delta_{\rm eff}$, see fig.\
\ref{pointers}, right. In order not to lose the information to be
measured, the {\em maximum} relaxation rate,  $\Gamma_{\rm r, max}$
reached at $\eta=1/2$ (i.e.\ $q=q_{\rm c}=\sqrt{\hbar/2M\Omega}$)
should be slow enough, such that by the time $\tau_{\rm ent}$ it takes
to ramp above $q_{\rm c}$, the information is not lost. In practice,
this can be achieved by  switching $q$ very fast, at a time $\tau_{\rm
ent}\ll\Gamma^{-1}_{\rm r, max}$,  to $q_{\rm c}$ and  slower afterwards,
when the actual measurement occurs.

In the ohmic spin-boson model \cite{Leggett,WeissBuch}, i.e.\  for
$J_{\rm eff}=2\pi\alpha\omega e^{-\omega/\omega_c}$,   a scaling of
$\Delta_{\rm eff}$ to zero   can be achieved through a dissipative
phase transition  at strong coupling to the bath
($\alpha>1$)\cite{Leggett,Zimanyi,Schmid,Helsinki}. This transition is
driven by the  entanglement with a {\em collective} state involving
the whole oscillator  bath. Ramping $\alpha$  to large values
increases $J_{\rm eff}(\omega)$ at {\em all} frequencies,  which leads to  rapid
relaxation {\em before} the scaling is established. Moreover,  it is
not known, how long it will take for the system to go through this
phase transition. Here, according to the adiabatic theorem, this time
is set through the inverse level spacing of the coupled system,  which
is infinite for the dense Ohmic spectrum. On the contrary, the model
studied in the present paper provides strong scaling of $\Delta_{\rm
eff}$ with predictably {\em slow} relaxation and gives a clear
prediction for the time scale of the entanglement set by the finite
level spacing.

\begin{figure}
\centerline{\psfig{figure=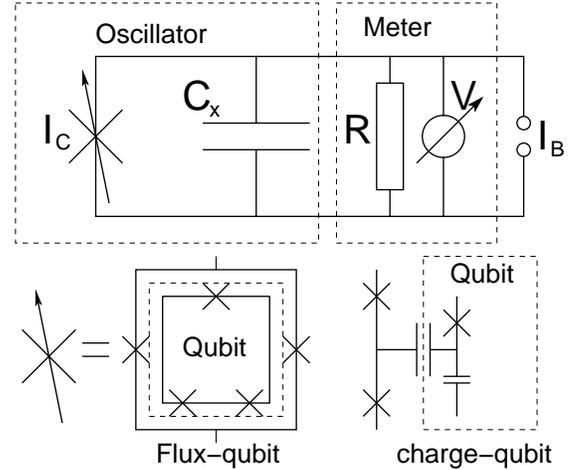,width=0.85\columnwidth}}
\caption{Underdamped read-out devices for superconducting flux (left)
and charge (right) quantum bits \label{tunJJ}.}
\end{figure}

This model does not generally predict the efficiency of the detection
In order to do so, I chose  a specific realization of the model, a
superconducting quantum bit \cite{Makhlin,HansScience}.  In this case,
the read-out device is a Josephson-junction, whose critical current
$I_{\rm 0}$  is influenced by the state of the qubit, either a
superconducting single-electron transistor \cite{Makhlin} or a
DC-SQUID  \cite{HansScience}, see Fig.\ref{tunJJ}. We study the
junction on  the superconducting branch at low bias current $I_{\rm
B}$.  We assume the  tunable junction to be shunted only by a very
large resistor $R$ and an external capacitance $C_{\rm x}$ and
consequently underdamped. This system has been studied in the weak damping
regime in Ref.\ \cite{EPJB} and resembles the one experimentally used in \cite{Caspar}.  The oscillator in our model represents the
plasma resonance \cite{tinkham}  of the tunable junction at
$\Omega=\sqrt{2eI_0/\hbar C_{\rm x}}(1-I_{\rm B}^2/I_{\rm
0}^2)^{1/4}$.   We can identify $M=C_{\rm x}(\hbar/2e)^2$,  $X=\phi$
(the Josephson phase) and  $q=\frac{\delta
I_c}{I_c}\frac{I_B}{\sqrt{I_0^2-I_B^2}}$,  where $\delta I_c/I_c$ is
the difference in critical current induced by the two fundamental
states of the qubit. The damping is provided  by the resistor and
leads to $\alpha=h q^2/2e^2 R =q^2\cdot 11.8 k\Omega/R$, and the
scaling exponent reads $\eta=q^2 \sqrt{C_{\rm x}I_c\hbar/8e^3}$.
When ramping $I_{\rm B}$, the junction  switches to a finite voltage
at $I_{\rm sw}<I_{\rm 0}$, which provides a measure for $I_{\rm 0}$.
This switching is a stochastic  process, so if the measurement is
repeated,  one finds a histogram of switching currents
\cite{VossWebb,Martinis} centered around $I_{\rm sw,0}$, whose width
$\delta I_{\rm sw}$  limits the resolution of this detector. In our
case, the switching  is predominantly due to thermal activation, where
we can express  $I_{\rm sw, 0}/I_{\rm
0}=1-(\log(\omega_T/\Gamma_S)/u_0)^{2/3}$ and $\delta I /I_{\rm
0}=(u_0^2\log(\omega_T/\Gamma_S))^{-1/3}$ through the dimensionless
height of the barrier at zero bias $u_0= (4\sqrt{2}/3)(\hbar
I_c/2ekT)$, the activation frequency $\omega_{\rm T}=2\Omega/\pi$ and
the ramp rate $\Gamma_s=d(q/q_{\rm max})/dt$.  The current  can be
switched within a time $\tau_{\rm sw}=\Omega^{-1}$, i.e.  the ramp
rate is limited by $\Gamma_{\rm s, max}=\Omega$.

In a flux qubit, one can realize  \cite{Caspar} $I_{\rm c}= 1\mu A$,
shunt with $C_{\rm x}=100pF$ and $R=10k\Omega$ and  $q_{\rm max}=0.05$
at a typical switching current level.  We will assume $\Delta_0=1$ GHz
and  $\epsilon=1$ GHz for the qubit.  These parameters are accessible
by  doubling the size of the sample studied in \onlinecite{Caspar}.
This leads to  $\Omega=2 GHz$, $\alpha=0.003$  and $\eta_{\rm
max}=3.5$, i.e.  $\Delta_{\rm eff}=0.03 \Delta_0$.  Entanglement sets
in at $q_{\rm c}=0.015$, where the  relaxation time is $\tau_{\rm r,
min}= \Gamma_{\rm r, max}^{-1}=5 \mu s$.  For  $1\%$  error, 
the first switch over this point has to be done $700 ns$, which
is way above $\Gamma_{\rm s, max}=500 ps$ and the adiabatic condition
$qdq/dt<(500 ns\;)^{-1}$ .   Close to the measuring point $I_{\rm
sw,0}$, we find  $\tau_{\rm r}=120 \mu s$ and $\tau_{\rm \phi}=100
ns$, which leaves a  huge measurement window.

For definiteness, we set the temperature to $T=200mK$,  we find, using
$\Gamma_s=(15 \mu s)^{-1}$, that $I_{\rm sw, 0}/I_{\rm 0}=0.96$,  and
$\delta I /I_{\rm 0}=0.35\%$,  so, because $q=5\%$, we have a
signal/noise ratio of about 14.  Hence, a single-shot
von-Neumann-measurement appears to be feasible within a gradual
improvement of technology.

For the readout for a charge qubit \cite{Nakamura} by a
superconducting single-electron-transistor (S-SET),one can  achieve
values of $q=0.5$.  within a charging energy $E_{\rm C, SET}=2 K$,
corresponding to a capacitance scale of $C=1fF$.  We take the
critical current of the SET  to be $I_{\rm c}=10 nA$ and a shunt of
$R=10k\Omega$ and $C_{\rm x}=1pF$  shunt capacitance. This leads to
$\eta=3.5$, $\Omega=2$ GHz and damping $\alpha=0.25$.  Assuming
$\Delta=1$ GHz and $\epsilon=1$ GHz, we find $\tau_{\rm R, min}=60
ns$, so  for $1\%$ error we have to go switch to $q_c$ in about $10
ns$, which is close to the limit of $\Gamma_{\rm s, max}=500 ps$.
however, may pose some challenge for the limiting time scales which
are {\em not} due to the on-chip circuitry. For the read-out step, we
find  $\tau_{\rm R}=25 \mu s$ and  $\tau_\phi=15 ns$. Applying the
histogram theory as above at $T=200 mK$ and $\Gamma_{\rm S}=(3 \mu
s)^{-1}$,  we end up with $I_{\rm sw, 0}/I_0=40\%$ and $\delta I_{\rm
sw}/I_{\rm c}=8\%$,  which can resolve our large signal of $q=30\%$ at
signal/noise of 4.  It has been shown \cite{Devoret,Saclay}, that
experimentally SETs {\em can} reach signal-to-noise  figures
comparable to the quantum limit, hence even though the resolution is
slightly less favorable then above, von-Neumann-measurements appear to
be possible. For qubits \cite{Vion}
operating in the regime of $E_J/E_c\simeq1$, 
more favorable parameters should be accessible.

The read-out of the detector by switching is only one, and not
necessarily the optimum alternative. Measurements could also be
performed by  detecting the kinetic inductance using the same
parameters.

A similar circuit, a {\em normal conducting}  SET with{\em out} the
shunt capacitor  has been thoroughly studied in Refs.\
\onlinecite{Makhlin,Shnirman}. In that case, the measurement is started by
rapidly switching the gate and monitoring the current.  It has been
shown, that in this way weak as well as strong  measurements can be
performed.  As a consequence of the direct coupling of a dense
spectrum of normal electrons to the qubit, the entanglement and the
decoherence are not as strongly separated as in our case.  Typical
\cite{Makhlin} mixing times during the whole measurement are  of the
order of $1\mu s$, i.e. one has to be able to monitor the current
through the SET on the scale of $100 ns$.  In our case, one has to
make the first entanglement  switch on a similar scale, but  has to be
monitor the voltage only afterwards, when mixing times are on the
scale of $10 \mu s$. These numbers clearly indicate an advantage of
the entanglement with the  intermediate oscillator.

I have proposed a strategy for performing detector-dominated
von-Neumann measurements on qubits, using entanglement with
coherent states of an harmonic oscillator. This system has been
quantitatively analyzed using   the spin-boson model and it has been
shown, that it has very favorable coherence and relaxation
properties. A connection between the familiar scaling of  the tunnel
matrix element and the degree of entanglement with the environment has
been established.  Realistic superconducting circuitry, which could
perform such  measurements within present-day technology, has been
proposed.

I acknowledge clarifying  discussions with J.\ von Delft, M.\
Governale,  M.\ Grifoni,  A.C.J.\ ter Haar, P.\ Hadley, P.\ Hakonen,
C.J.P.M.\ Harmans, S.\ Kleff, L.\ Levitov, S.\ Lloyd, A.\ Lupascu,
J.E.\ Mooij, T.P.\ Orlando, A.\ Shnirman, and  C.H.\ van der Wal, as
well as support by the EU through TMR Supnan and Quiprocone and
through ARO under contract Nr.\ P-43385-PH-QC.

%\twocolumn[\hsize\textwidth\columnwidth\hsize\csname@twocolumnfalse\endcsname
%\input{zeitpfeil.latex}

%]

\begin{thebibliography}{10}
\bibitem{FN} This non-separable state is not fully entangled, the
degree of entanglement is $1-e^{-\eta}$
\end{thebibliography}
\end{document}